\documentclass[twocolumn,prl,superscriptaddress]{revtex4}
\usepackage{amssymb}
\usepackage{amsfonts}
\usepackage{amsmath}
\usepackage{graphicx}
\usepackage{graphics}
\usepackage{bm}
\usepackage{mathdots}
\usepackage{calc}
\usepackage{dcolumn}
\newcommand{\mX}{{\mathcal X}}
\newcommand{\mZ}{{\mathcal Z}}

\newcommand{\mO}{{\mathcal O}}
\newcommand{\mG}{{\mathcal G}}

\usepackage{pstricks}
\usepackage{pst-plot}
\begin{document}
\title{Quantum pigeonhole effect, Cheshire cat and contextuality}

\author{Sixia Yu}
\affiliation{Centre for Quantum Technologies, National University of Singapore, 2 Science Drive 3, Singapore 117542}
\author{C.H. Oh}
\affiliation{Centre for Quantum Technologies, National University of Singapore, 2 Science Drive 3, Singapore 117542}
\affiliation{Physics Department, National University of Singapore, 2 Science Drive 3, Singapore 117542}
\begin{abstract}
A kind of paradoxical effects has been demonstrated that the pigeonhole principle, i.e., if three pigeons are put in two pigeonholes then at least two pigeons must stay in the same hole, fails in certain quantum mechanical scenario. Here we shall show how to associate a proof of Kochen-Specker theorem with a quantum pigeonhole effect and vise versa, e.g., from state-independent proofs of Kochen-Specker theorem some kind of state-independent quantum pigeonhole effects can be demonstrated. In particular, a state-independent version of the quantum  Cheshire cat, which can be rendered as a kind of quantum pigeonhole effect about the trouble of putting two pigeons in two pigeonholes, arises from Peres-Mermin's magic square proof of contextuality.
\end{abstract}
\maketitle

Quantum theory confronts us with a reality that is radically different from a classical one because of its contextuality. Roughly speaking, quantum contextuality refers to the property that any realistic theory trying to complete quantum mechanics so as to avoid indeterministic measurement  outcomes has to be contextual, i.e., the predetermined outcomes of measuring an observable may depend on which set of compatible observables that might be measured along side. This is a seminal no-go theorem proved by Kochen and Specker (KS) \cite{ks}  and independently by Bell \cite{bell2}. Bell's nonlocality as revealed by, e.g., the violation of Bell's inequality \cite{bell,bi} or arguments without inequality \cite{ghz,mermin,tang,hardy}, is a special form of contextuality enforced by space-like separation. In different scenarios there are different proofs of Kochen-Specker theorem, which can be state-dependent \cite{clif,pent}, state-independent and deterministic \cite{ks,cab,peres}, and statistical yet state-independent \cite{yo}.

Recently a kind of quantum pigeonhole effect \cite{qpe,vaid} was demonstrated that the pigeonhole principle, which states that if three pigeons are to be put into two pigeonholes then two of the pigeons must stay in the same hole, does not hold in some pre- and post-selection scenarios. This effect is even promoted to be a principle to explore the nature of the quantum correlations. Consider a system of three qubits representing three pigeons in two pigeonholes, i.e., two eigenstates $\{|0\rangle,|1\rangle\}$ of $Z$, where three Pauli matrices and identity matrix are denoted simply  by $\{X,Y,Z,I\}$.
Initially the system is prepared in the state  $|\psi_i\rangle=|+,+,+\rangle$, which is the common $+1$  eigenstate of commuting observables $\{X_1,X_2,X_3\}$. At the final stage, $y$ component of each qubit is measured $\{Y_1,Y_2,Y_3\}$  and only those outcomes with three +1 are kept, i.e., the common $+1$ eigenstate $|\psi_f\rangle=|0,0,0\rangle_Y$ is post-selected. In the intermediate stage between the preparation and post-selection, one asks what would happen if we had tested whether each pair of two qubits is in the same state or not. This is equivalent to measure the observable $Z_{ab}=Z_aZ_b$ on each pair of qubits $a$ and $b$ since the outcome $+1$ or $-1$, which corresponds to the projectors $\Pi_{ab}^{\pm}=(I_{ab}\pm Z_{ab})/2$, indicates that two qubits are in the same or different states, respectively, with respect to the the computational basis. It was then argued that  the detectors corresponding to $\{\Pi_{ab}^+\}$ would never fire because of the identities
\begin{equation}\label{cd}
\langle\psi_i|\Pi_{ab}^+|\psi_f\rangle=0 \quad (a,b=1,2,3).
\end{equation}
 As a result the detector corresponding to $\Pi_{ab}^-$ would always fire for each pair of qubits $a,b=1,2,3$, meaning that each pair of qubits would have stayed in different states, a violation to the pigeonhole principle.

The quantum pigeonhole effect described above is obviously a kind of pre- and post-selection paradox and, as pointed out by Leifer and Spekkens \cite{ls},  any such kind of paradox is associated with a proof of quantum contextuality. In any noncontextual realistic model, according to Kochen and Specker \cite{ks}, there is a so called KS value assignment of $\{0,1\}$ to all the rays, i.e., rank-1 projections, in the relevant Hilbert space satisfying the following three rules. The {\it noncontextuality rule} states that a ray is assigned to value 0 or 1 regardless of which complete orthonormal bases it belongs to, the {\it orthogonality rule} states that two orthogonal rays cannot be assigned to value 1 simultaneously, and the {\it completeness rule} states that within a complete basis orthonormal basis there is at least one ray  that is assigned to value 1.
A finite set of rays having no KS value assignment is a proof of the KS theorem, i.e., a demonstration of quantum contextuality, and the proof is state-dependent if some states are assigned to value 1 {\it a priori}. The absence of  KS value assignments is sufficient for the nonexistence of a noncontexual model but not necessary. There are also state-independent proofs admitting KS value assignments \cite{yo}.

In the above demonstration of quantum pigeonhole effect, by denoting  $|\Phi_\pm\rangle\propto{ |00\rangle\pm|11\rangle}$ and $|\Psi_\pm\rangle\propto |01\rangle\pm|10\rangle$, the set of all the  relevant 34 pure states
\begin{equation}
|\psi_i\rangle,|\psi_f\rangle, \{|\Phi_\pm\rangle_{ab}|\mu\rangle_c,|\Psi_\pm\rangle_{ab}|\mu\rangle_c\}, \{|\mu,\nu,\tau\rangle\},
\end{equation}
where
$(a,b,c)$ denotes one of possible cyclic permutations of $(1,2,3)$ and $\mu,\nu,\tau=0,1$, provides a state-dependent proof of KS theorem if $|\psi_i\rangle$ and $|\psi_f\rangle$ are assigned to value 1. In fact
for any possible choice of $(a,b,c)$ and $\mu=0,1$, both two rays $|\Phi_\pm\rangle_{ab}|\mu\rangle_c$   have to be assigned to value 0 because of Eq.(\ref{cd}) and the orthogonality rule. As a result, either $|\Psi_+\rangle_{ab}|\mu\rangle_c$ or $|\Psi_-\rangle_{ab}|\mu\rangle_c$ has to be assigned to value 1 for any given $\mu=0,1$ and $a,b$, according to the completeness rule. Or equivalently, all three rank-2 projections $\{\Pi_{ab}^-\}$ must be assigned to value 1.
Due to the pigeonhole principle, in each one of the eight computational bases $\{|\mu,\nu,\tau\rangle\}$  at least two qubits are in the same states and therefore is orthogonal at least to one of the three subspaces with projections $\{\Pi_{ab}^-\}$. That is to say all eight computational bases have to be assigned to value 0, which contradicts the completeness rule.

All
three rules of KS value assignment are also assumed implicitly or explicitly in the demonstration of the quantum pigeonhole effect.  The orthogonality and completeness rules, which can be enforced by Aharonov-Bergmann-Lebowitz rule \cite{abl} for pre- and post-selection, have been used in the arguments against certain outcomes of the intermediate measurements. The noncontextuality has also been assumed implicitly in the intermediate measurement of the observable $Z_{ab}$ for each pair of qubits, which appears in two different measurement contexts in order to extract a contradiction. The first measurement context is given by $\{X_{ab},Y_{ab}\}$ which has been used to show that the detector $\Pi_{ab}^+$ would never fire, i.e., $Z_{ab}$ would take value $-1$ indication qubit $a$ and $b$ are in different states. The second measurement context is given by $\{Z_1,Z_2,Z_3\}$ whose eigenstates represent all possible configurations of three pigeons in two pigeonholes. There is a contradiction only if the value of $Z_{ab}$ obtained in the first context is used in the second context. It is clear that no paradox will be present if the assumption of noncontextuality is dropped. The counterfactual measurements might yield contextual values.

For other possible outcomes of the post-selection measurement, similar quantum pigeonhole effects can still be demonstrated \cite{qpe}. Here we shall remove further the dependency of the initial state so that we have a state-independent version of the quantum pigeonhole effect. On a system of three qubits we at first measure the complete set $\{X_1,X_2,X_3\}$ with outcomes $s_1,s_2,s_3=\pm1$ being arbitrary, and at the final stage we measure the complete set $\{Y_1,Y_2,Y_3\}$ with outcomes  $t_1,t_2,t_3=\pm1$  being arbitrary. By the first measurement the system is prepared in the state $|\psi_i^s\rangle=|s_1,s_2,s_3\rangle$ and  by the second measurement the system is post-selected into the state $|\psi_f^t\rangle=|t_1,t_2,t_3\rangle_Y$. Let us now examine what would happen if we had measured $Z_{ab}$ at the intermediate stage between the preparation and post-selection. Denoting $v_{ab}=s_as_bt_at_b$ we have
\begin{eqnarray}\label{test}
v_{ab}\langle\psi_i^s|\Pi_{ab}^{v_{ab}} |\psi^t_f\rangle&=&\langle\psi_i^s|X_{ab}\Pi_{ab}^{v_{ab}} Y_{ab}|\psi^t_f\rangle\nonumber\\
&=&-\langle\psi_i^s|X_{ab}\Pi_{ab}^{v_{ab}} Z_{ab}X_{ab}|\psi^t_f\rangle\nonumber\\
&=&-{v_{ab}} \langle\psi_i^s|\Pi_{ab}^{v_{ab}} |\psi^t_f\rangle=0
\end{eqnarray}
for arbitrary $a,b=1,2,3$. That is, the detector $\Pi_{ab}^-$ (or $\Pi_{ab}^+$) would never fire if $v_{ab}=-1$ (or 1) so that qubits $a$ and $b$ would have stayed in the same (or different) states, respectively. Since
$ v_{12}v_{23}v_{13}=1$ there is an even number of pairs of qubits such that $v_{ab}=-1$, i.e., among three qubits there is an even number of pairs of qubits that would have stayed in the same state. This is impossible due to the fact that to put three pigeons into two pigeonholes the number of pairs of pigeons staying in the different holes must be even.

\begin{figure}
\includegraphics[scale=1]{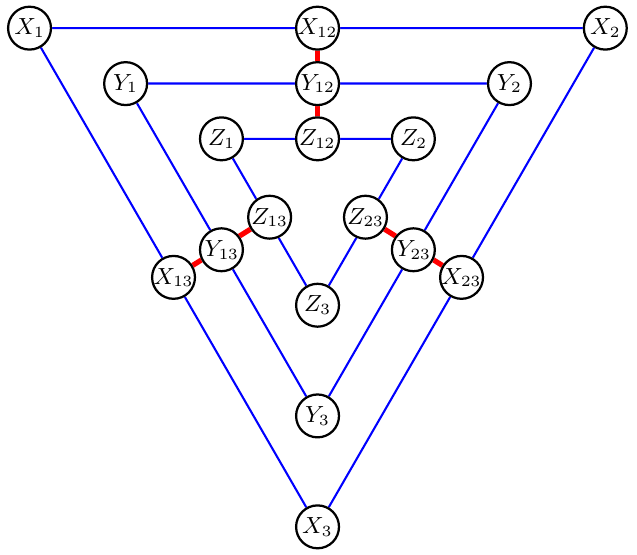}
\caption{(Color online.) Waegell-Aravind's triangle configuration for a proof of Kochen-Specker theorem in the case of three qubits. }
\end{figure}

The above state-independent version of the quantum pigeonhole paradox also gives rise to a state-independent proof of Kochen-Specker theorem due to Waegell and Aravind \cite{wa}. This proof includes  48 rays defined by eigenstates of 6 maximal sets of mutually commuting observables
$\{X_a\},\{Y_a\},\{Z_a\},\{X_{ab},Y_{ab},Z_c\},
$ with $(a,b,c)=(1,2,3)$. There is an elegant parity proof given by the Waegell-Aravind's triangle configuration as shown in Fig.1. Quantum mechanically, the product of three observables connected by each thin (blue) straight line is 1 while the product of three observables connected by each thick (red) line is $-1$. Thus the product of all those line products of three observables equals $-1$. In a noncontextual realistic theory, where all  observables have some predetermined values $\pm1$ independent of the contexts, the same product yields value 1 since each observable appears twice in the product.

Any proof of Kochen-Specker theorem via logical contradiction can give rise to the demonstration of some sort of  quantum pigeon effect.  This is because in each KS proof there is at least a basis, one of which can be taken without loss of generality to be the computational basis. The logical contradiction can be pushed back to the impossibility of the KS value assignment to this computational basis. We note that the computational bases represent all possible configurations of pigeons in holes and satisfy obviously the orthogonality and completeness rules, i.e., one and only one configuration is realized. Thus the contradiction as seen in a KS proof can thus be regarded as a violation of some kind of generalized pigeonhole principle. For example the 3-box paradox \cite{3box}, which is associated with Clifton's state-dependent proof of KS theorem \cite{clif}, roughly shows that a single pigeon, when put into three pigeonholes, can stay simultaneously in two holes. Now let us see some more examples.

As the first example, some kind of robust quantum pigeonhole effect can be demonstrated by using  Peres-Mermin's magic square for three qubits \cite{wa} as shown in Table I (left). Let the system be prepared in any state $|\psi_i\rangle$ (may even be mixed) in the common $+1$ eigenspace  of three observables  $\{X_{12},X_{23},X_{13}\}$ in the fist row, which is spanned by $\{|\pm,\pm,\pm\rangle\}$. At the final stage we make a post-selection to any state $|\psi_f\rangle$ in the common +1 eigenspace of three observables  $\{Y_{12},Y_{23},Y_{13}\}$ in the third row, which is spanned by $\{|\pm,\pm,\pm\rangle_Y\}$.  At the immediate stage we ask what would happen if we had measured three observables $\{Z_{12},Z_{23},Z_{13}\}$ in the second row, testing whether each pair of qubits was in the same state or not. In this general setting  Eq.(\ref{cd}) still hold so that each pair of qubits would have stayed in different states, a violation to the pigeonhole principle. The dependency of the pre- and post-selection can be easily removed and the generalization of Peres-Mermin's magic square as well as the quantum pigeonhole effect to the case of an odd number of qubits is straightforward.

\begin{table}
$$
\begin{array}{|c|c|c|}\hline
X_1X_2& X_2X_3& X_1X_3\\
\hline
Z_1Z_2& Z_2Z_3 & Z_1Z_3\\
\hline
Y_1Y_2& Y_2Y_3& Y_1Y_3\\
\hline
\end{array}
\quad\quad
\begin{array}{|c|c|c|}\hline
X_1& X_2& X_1 X_2\\
\hline
Z_2 & Z_1 & Z_1 Z_2\\
\hline
X_1Z_2 & Z_1X_2 & Y_1Y_2\\
\hline
\end{array}
$$
\caption{Peres-Mermin's magic square for three qubits (left) and for two qubits (right). }
\end{table}

As the second example, a kind of paradoxical effect can be demonstrated in the case of two pigeons in two pigeonholes, which turns out to be the quantum Cheshire cat \cite{cat}, a curious situation of a grin without a cat encountered by Alice only in wonderland and  demonstrated recently in a neutron experiment \cite{catn}. Consider a spin half particle, e.g., neutron, in an interferometer with two paths, where the cat is represented by the particle and its grin by its spin. This is effectively a two-qubit system with the first qubit representing the spatial degree of freedom, i.e., the path, while the second qubit representing its spin.
Let the system be prepared initially in the state $|\psi_i\rangle=|\Phi_+\rangle$ and post-selected to $|\psi_f\rangle=|+\rangle_1|0\rangle_2$ at the final stage. Because $\langle\psi_i |\Pi_1^{Z_1}|\psi_f\rangle=0$ the particle would take path $|0\rangle_1$ if we had observed its path, where we have denoted  by $\Pi^O_u$ the projection to the eigenspace of observable $O$ corresponding to eigenvalue $(-1)^u$ with $u=0,1$. Because $\langle\psi_i |\Pi_1^{X_2}|\psi_f\rangle=0$ if we had observed its spin components along $x$ direction the answer would be spin up $|+\rangle_2$. Because $\langle\psi_i |\Pi^{Z_1X_2}_0|\psi_f\rangle=0$ we see that the path and the spin are anti-correlated, i.e., if the spin is up $|+\rangle_2$ then the path $|1\rangle_1 $ should be taken while if the spin is down $|-\rangle_2$ then the path $|0\rangle_1$ should be taken. To sum up, given the pre and post-selection, the particle would surely travel along path $|0\rangle$ and its spin would be up along $x$ direction while the path and spin would be anti-correlated, i.e., its spin would be measured along a different path $|1\rangle$ from what is  taken by the particle, i.e., $|0\rangle$, which is exactly our quantum Cheshire cat.
This demonstration of quantum Cheshire cat, being a kind of pre- and post-selection paradox,  also gives rise to a state-dependent proof of Kochen-Specker theorem including the eigenstates of $\{Z_1,Z_2\}$,
$\{X_1,X_2\}$ and $\{Z_1X_2,X_1Z_2\}$ in addition to $|\psi_i\rangle$ and $|\psi_f\rangle$.

By considering suitable pre- and post-selection measurements the dependency of the quantum Cheshire cat on the preparation and post-selection can be removed. For the preparation we measure observables $\{X_{12},Z_{12}\}$ while for the post-selection we measure observables $\{X_1,Z_2\}$. Thus initially the system is prepared in the state $|\psi_i\rangle=|\Phi_{\alpha\beta}\rangle$ which is the common eigenstate of  $X_{12}$ and $Z_{12}$ corresponding to eigenvalues $(-1)^\alpha$ and $(-1)^\beta$, respectively, with $\alpha,\beta=0,1$, and post-selected into the state $|\psi_f\rangle=|(-1)^\mu\rangle_1|\nu\rangle_2$ with $\mu,\nu=0,1$. The following three observations constitute the quantum Cheshire cat. Firstly, if we had observed which path were taken by the particle we would find the particle in the path $|u\rangle_1 $ with $(-1)^u=(-1)^{\beta+\nu}$ since
\begin{eqnarray*}
(-1)^{\beta+\nu}\langle\psi_i |\Pi^{Z_1}_u|\psi_f\rangle&=&\langle\psi_i| Z_{12}\Pi^{Z_1}_uZ_2|\psi_f\rangle\\&=&(-1)^u\langle\psi_i |\Pi^{Z_1}_u|\psi_f\rangle
\end{eqnarray*}
vanishes otherwise.
Secondly, if we had observed the spin along $x$ direction we would find the spin in the state $|(-1)^v\rangle_2$ with $(-1)^v=(-1)^{\alpha+\mu}$ since
\begin{eqnarray*}
(-1)^{\alpha+\mu}\langle\psi_i |\Pi^{X_2}_v|\psi_f\rangle&=&\langle\psi_i| X_{12}\Pi^{X_2}_vX_1|\psi_f\rangle\\&=&(-1)^v\langle\psi_i |\Pi^{X_2}_v|\psi_f\rangle
\end{eqnarray*}
vanishes otherwise. Lastly, if we had observed the correlation $\{Z_1X_2\}$ then the detector $\Pi_w^{ZX}$ would never fire if $(-1)^w=(-1)^{\alpha+\beta+\mu+\nu}$ because in this case
\begin{eqnarray*}
(-1)^{\alpha+\beta+\mu+\nu}\langle\psi_i |\Pi^{ZX}_w|\psi_f\rangle &=&-\langle\psi_i| Y_{12}\Pi^{ZX}_wX_1Z_2|\psi_f\rangle\\&=&-(-1)^w\langle\psi_i |\Pi^{ZX}_w|\psi_f\rangle
\end{eqnarray*}
vanishes. This means that the spin state $|(-1)^v\rangle_2$, as revealed by the measurement $\{X_2\}$, has to be correlated with  the path $|u+1\rangle_1$ which is different from the path $|u\rangle_1$ taken by the particle as revealed by the measurement $\{Z_1\}$. This state-independent version of the quantum Cheshire cat turns out to be exactly the parity proof by Peres-Mermin's magic square \cite{pm} as shown in Table I (right). Three commuting observables in three columns are taken as the post-selection, intermediate,  and preparation measurements respectively. If we take three sets of row observables instead then we obtain a paradoxical effect that it is impossible to put two pigeons in two holes in a pre- and post-selection scenario.

\begin{figure}
\includegraphics[scale=0.8]{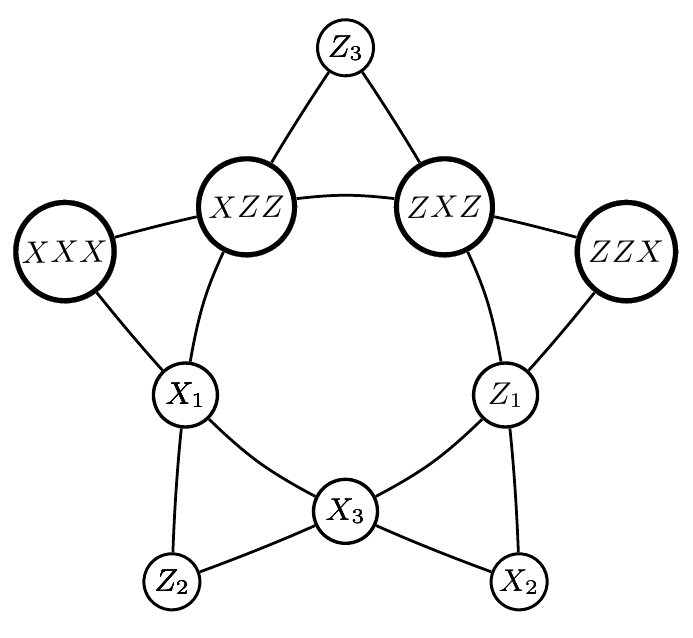}
\caption{Mermin's magic pentagram for
 three qubits. }
\end{figure}

As the last example, let us see how to derive pigeonhole effects from Greenberger-Horne-Zeilinger (GHZ) paradoxes \cite{ghz,mermin,tang}. Consider Mermin's magic pentagram proof of KS theorem for three qubits as shown in Fig.2. Let the system be prepared in the GHZ state $|\psi_i^s\rangle$, i.e., one of the common eigenstates of three commuting observables $\{G_a=X_aZ_bZ_c\}$, with $(a,b,c)$ being one of three cyclic permutations of $(1,2,3)$,  corresponding to eigenvalues $\{s_a=\pm1\}$ respectively. As the post-selection we measure observables $\{X_1,X_2,X_3\}$ with outcomes denoted by $t_a=\pm1$, i.e., the post-selection state is $|\psi_f^t\rangle=|t_1,t_2,t_3\rangle$. We note that outcomes must satisfy $st=-1$ where $s=s_1s_2s_3$ and $t=t_1t_2t_3$ since otherwise we will have orthogonal pre- and post-selected states $$-s\langle\psi_i^s|\psi_f^t\rangle=\langle\psi_i^s|X_{123}|\psi_f^t\rangle=t\langle\psi_i^s|\psi_f^t\rangle=0.$$ For each successful pre- and post-selection, if we had measured observables $Z_{ab}$ for $a,b=1,2,3$ then we would register outcome ${v_{ab}}=t_cs_c$ because otherwise
\begin{eqnarray*}
s_ct_c\langle\psi_i^s|\Pi^{v_{ab}}_{{ab}}|\psi_f^t\rangle&=&
\langle\psi_i^s|G_c\Pi^{v_{ab}}_{{ab}}X_c|\psi_f^t\rangle\\
&=&{v_{ab}}\langle\psi_i^s|\Pi^{v_{ab}}_{{ab}}|\psi_f^t\rangle
\end{eqnarray*}
would vanish. Since $v_{ab}=\pm1$ indicates whether two qubits are in the same state (computational basis) or not, from $v_{12}v_{23}v_{13}=st=-1$ it follows that there is an odd number of qubit pairs in different state, which is impossible because if we put three pigeons in two holes the number of pairs of pigeons in different state should be even. This is a state-independent version of an early proposal of the pigeonhole effect \cite{vaid} using entangled states. GHZ paradoxes for multi (even) levels systems and multi particles can be systematically constructed from the so called GHZ graph \cite{tang} and corresponding graph states. Each paradox will give rise to a similar quantum pigeonhole effect as above (see Appendix).

To summarize, we have established an intimate relation between the quantum pigeonhole effects and proofs of KS theorem. On the one hand this relation enables us to understand those paradoxical effects in terms of contextuality, since the assumption of noncontextuality is indispensable in all these paradoxical effects.  On the other hand this relation helps to obtain  from known proofs of KS theorem some other interesting effects, such as state-independent versions and the impossibility of putting two pigeons in more than two pigeonholes. Most interestingly, the latter effect is related to the quantum Cheshire cat, for which we also find a state-independent version with the help of Peres-Mermin's magic square. From the experimental point of view, the removal of the dependency on the post-selection states enhances the success probability by 300\%. Lastly, it will be interesting to find out what kinds of paradoxical effects correspond to those KS proofs without logical contradictions.

We are indebted to M. Howard for pointing out a mistake in previous version and L. Vaidman for providing a helpful reference \cite{vaid}. This work is funded by the Singapore Ministry of Education (partly through the Academic Research Fund Tier 3 MOE2012-T3-1-009) and the National Research
Foundation, Singapore (Grant No. WBS: R-710-000-008-271).


\section*{Appendix}

Here we shall demonstrate how to derive quantum pigeonhole paradoxes from GHZ paradoxes obtained from qudit graph states corresponding to GHZ graphs \cite{tang}.

A weighted graph is defined by a set $V=\{1,2,\ldots,n\}$ of $n$ vertices and a set of edges specified by the adjacency matrix $\Gamma$ whose entry $\Gamma_{ab}\in {\mathbb Z}_p=\{0,1,\ldots,d-1\}$ denotes the weight of the edge connecting two vertices $a,b$. We consider here only undirected graph without self loop so that the adjacent matrix is symmetric and has zero diagonal entries. A GHZ graph \cite{tang} is defined to be a weighted graph $G=(V,\Gamma)$ satisfying
$$d_a:=\sum_{b\in V}\Gamma_{ab}\equiv 0\mod d\quad (a\in V),$$
and
$$W:=\sum_{a>b}\Gamma_{ab}\not\equiv 0\mod d.$$
As a result $d$ must be even and $\omega^W=-1$ with $\omega=e^{i\frac{2\pi}d}$. GHZ graphs exist for all values of $n\ge3$ and even values of $d$ with some examples shown in Fig.3.

Consider a system of $n$ qudits labeled with $V$. For each qudit $a\in V$ we denote by $\{|k\rangle_a\mid {k\in {\mathbb Z}_d}\}$ its computational basis and by
\begin{equation*}
\mX_a=\sum_{k\in {\mathbb Z}_d}|k\rangle\langle k\oplus1|_a,\quad \mZ_a=\sum_{k\in {\mathbb Z}_d}\omega^{k}|k\rangle\langle k|_a
\end{equation*}
its generalized bit and phase shifts for which it holds $\mZ^d_a=\mX_a^d=I$ and the commutation rule $\mX_a\mZ_a=\omega\mZ_a\mX_a$. We denote
$$\mX_V:=\bigotimes_{a\in V}\mX_a,\quad \mZ_{N_a}:=\bigotimes_{b\in V}\mZ_b^{\Gamma_{ab}},\quad a\in V.$$
The graph state $|G\rangle$ for a given graph $G=(V,\Gamma)$ is defined by the $+1$ common eigenstate of $n$ commuting unitary observables $\{\mG_a=\mX_a\mZ_{N_a}\}$, i.e., $\mG_a|G\rangle=|G\rangle$ for all $a\in V$. For a GHZ graph it holds
$$\prod_{a\in V} \mG_a=\omega^W\mX_V\prod_{a\in V}\mZ_{N_a}=-\mX_V\bigotimes_{a\in V}\mZ_a^{d_a}=-\mX_V$$
based on which we can construct a GHZ paradox as well as a state-independent parity proof of KS theorem as shown in Table II.

\begin{figure}
\includegraphics[scale=1]{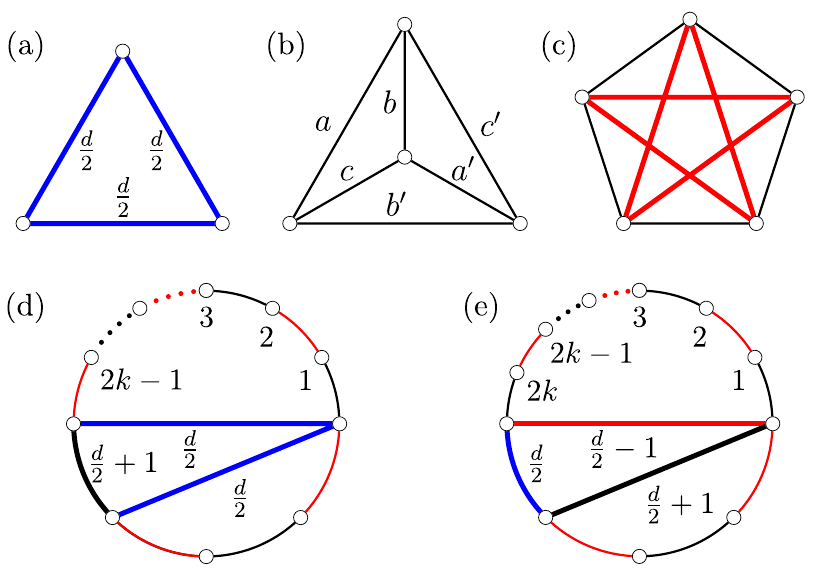}
\caption{(Color online \cite{tang}.) Some examples of GHZ graphs. Unlabeled thin black or red edges have weight 1 or $d-1$, respectively. Thick black and red edges have weight $d/2+1$ and $d/2-1$, respectively. All possible GHZ graphs on 3 and 4 vertices are shown in (a) and
(b), where $a^\prime=d/2+a$, $b^\prime=d/2+b$, and $c^\prime=d/2+c$ with $a+b+c=d/2$.}
\end{figure}

\begin{table}
$$
\begin{array}{|c|c|cc|c|c|c|}\hline
\mX_1& \mX_2&\ldots&\ldots& \mX_{n-1}&\mX_n& \mX_V^\dagger\\
\hline
\mZ_{N_1}& \mZ_{N_2}&\ldots&\ldots& \mZ_{N_{n-1}}&\mZ_{N_n}& I\\
\hline
\mG_1^\dagger& \mG_2^\dagger&\ldots&\ldots& \mG_{n-1}^\dagger&\mG_n^\dagger& \mX_V\\
\hline
\end{array}
$$
\caption{Mermin's magic configuration for qudits. }
\end{table}

All unitary observables $\{\mO_{rc}\}$ labeled with rows $r=1,2,3$ and columns $c=1,2,\ldots, n+1$ in Table II have spectrum ${\mathbb U}_d=\{\omega^k\mid k\in {\mathbb Z}_d\}$ and observables in the same column or row commute with each other.  Quantum mechanically, it is clear that we have an identity
\begin{equation*}
\prod_{r=1}^3\prod_{c=1}^{n+1}\mO_{rc}\prod_{c=1}^{n+1}\prod_{r=1}^3\mO_{rc}^\dagger=-1.
\end{equation*}
In noncontextual models $3(n+1)$ observables $\{\mO_{rc}\}$ assume predetermined noncontextual values $v_{rc}\in {\mathbb U}_d$ and the above product with observables replaced by their realistic values,  since $|v_{rc}|=1$, is equal to $1\not=-1$, a contradiction.

A quantum pigeonhole paradox about the trouble of putting $n$ pigeons into $d$ holes can be demonstrated as follows. Let the system of $n$ qudits be prepared in the common eigenstate $|\psi_i^g\rangle$ of $\{\mG_a\}$ corresponding to eigenvalues $\{g_a\in {\mathbb U}_d\}$ and be post-selected into the common eigenstate $|\psi_f^h\rangle$ of $\{\mX_a\}$ with outcomes $\{h_a\in {\mathbb U}_d\}$. Outcomes $\{h_a\}$ and $\{g_a\}$ are not possible unless  $$\prod_ah_a=-\prod_ag_a$$
because otherwise the pre- and post-selected states will be orthogonal. Given successful pre- and post-selection, if we had measured the observable $\mZ_{N_a}$ in the intermediate stage we would obtain outcome $S_a=g_ah_a^*$ for each $a\in V$ because otherwise
\begin{eqnarray*}
S_a\langle\psi_i^g| \Pi^{\mZ_{N_a}}_{S_a}|\psi_f^h\rangle&=&\langle\psi_i^g|\mZ_{N_a} \Pi^{\mZ_{N_a}}_{S_a}|\psi_f^h\rangle\\
&=&\langle\psi_i^g|\mG_a \Pi^{\mZ_{N_a}}_{S_a} \mX_a^\dagger|\psi_f^h\rangle\\
&=&g_ah_a^*\langle\psi_i^g| \Pi^{\mZ_{N_a}}_{S_a} |\psi_f^h\rangle
\end{eqnarray*}
will vanish, where we have denoted by $\Pi_O^{\mO}$ the projection to the eigenspace of $\mO$ corresponding to eigenvalue $O$. However if these values $\{S_a\}$ are measured in the context $\{\mZ_a\}$, whose outcomes are denoted by $\{s_a\in{\mathbb U}_d\}$, or a possible configuration of $n$ pigeons in $d$ pigeonholes, then we should have $\prod_aS_a=\prod_as_a^{d_a}=1$, a contradiction. In other words, the possible answers to $n$ questions about how $n$ pigeons are distributed in $d$ pigeonholes, namely, $\mZ_{N_a}$ with $a\in V$, provided by measurement context $\{\mZ_{N_a},\mX_V\}$ are incompatible with every configuration of directly putting $n$ pigeons into $d$ levels, namely, the measurement context $\{\mZ_a\}$.

To demonstrate a paradox, the pre-selected state is not necessarily an entangled state and can also be chosen to be a product state. For preparation we measure observables $\{\mZ_a\}$ with outcomes denoted by $s_a$, respectively, and for post-selection we still measure observables $\{\mX_a\}$ with outcomes $h_a$. Thus the pre-selected state $|\psi_i^s\rangle$ is the common eigenstate of $\{\mZ_a\}$ corresponding to eigenvalue $\{s_a\}$ while the post-selected state $|\psi_f^h\rangle$ is the common eigenstate of $\{\mX_a\}$ corresponding to eigenvalue $\{h_a\}$. All outcomes are possible. In the intermediate stage, if we had measured observable $\mG_a$ then we would obtain outcome $g_a=h_aS_a$ where $S_a=\prod_bs_b^{\Gamma_{ab}}$ since otherwise
\begin{eqnarray*}
h_aS_a\langle\psi_i^s| \Pi^{\mG_a}_{g_a}|\psi_f^h\rangle&=&\langle\psi_i^s|\mZ_{N_a} \Pi^{\mG_a}_{g_a} \mX_a|\psi_f^h\rangle\\
&=&{g_a}\langle\psi_i^s| \Pi^{\mG_a}_{g_a}|\psi_f^h\rangle
\end{eqnarray*}
would vanish. The dilemma lies in the fact that on the one hand the outcome for measuring observable $\mX_V$ should be $-\prod_ag_a=-\prod_a h_a$, since $\prod_aS_a=1$, from the above arguments and on the other hand $|\psi_f^h\rangle$ is an eigenstate of $\mX_V$ corresponding to eigenvalue $\prod_ah_a$.

\newpage
{\mbox{$\quad$}}
\newpage
\end{document}